\documentclass[aps,prx,reprint,groupedaddress,superscriptaddress,nofootinbib,twocolumn]{revtex4-1}
\usepackage{graphicx}
\usepackage{amsmath}
\usepackage{fancyhdr}
\usepackage{epsfig}
\usepackage{esvect}
\usepackage{cancel}
\usepackage{color}
\usepackage{epstopdf}
\usepackage{subcaption}
\usepackage{bm}

\usepackage{comment}
\setcitestyle{square}

\begin{document}
\title{The refractory period matters:\\
unifying mechanisms of macroscopic brain waves}

\author{Corey Weistuch}
\affiliation{Laufer Center for Physical and Quantitative Biology, Stony Brook University}
\affiliation{Department of Applied Mathematics and Statistics, Stony Brook University}

\author{Lilianne R. Mujica-Parodi}
\affiliation{Laufer Center for Physical and Quantitative Biology, Stony Brook University}
\affiliation{Department of Biomedical Engineering, Stony Brook University}
\affiliation{Department of Physics and Astronomy, Stony Brook University}
\affiliation{Program in Neuroscience, Stony Brook University}
\affiliation{Athinoula A. Martinos Center for Biomedical Imaging, Massachusetts General Hospital and Harvard Medical School}

\author{Ken Dill}
\affiliation{Laufer Center for Physical and Quantitative Biology, Stony Brook University}
\affiliation{Department of Physics and Astronomy, Stony Brook University}
\affiliation{Department of Chemistry, Stony Brook University}

\begin{abstract}
The relationship between complex, brain oscillations and the dynamics of individual neurons is poorly understood.  Here we utilize Maximum Caliber, a dynamical inference principle, to build a minimal, yet general model of the collective (mean-field) dynamics of large populations of neurons.  In agreement with previous experimental observations, we describe a simple, testable mechanism, involving only a single type of neuron, by which many of these complex oscillatory patterns may emerge.  Our model predicts that the refractory period of neurons, which has been previously neglected, is essential for these behaviors.
\end{abstract}

\date{\today}

\maketitle

\section*{Introduction} 

A major interest in neuroscience is understanding how macroscopic brain functions, such as cognition and memory, are encoded at the microscale of neurons and their topological connectivities.  One of the significant developments in this direction was the Wilson-Cowan (WC) model, describing the averaged behavior of large populations of simple excitatory and inhibitory neurons in terms of a set of coupled, mesoscale differential equations \cite{wilson1972excitatory,wilson1973mathematical,destexhe2009wilson}.  With only a few physical parameters, WC provided one of the first mechanisms for simple (single-frequency) oscillations across the brain, such as the hyper-synchronized dynamics observed during epileptic seizures \cite{destexhe2009wilson,shusterman2008baseline}.  More recently, generalized WC-like models have been used to describe heterogeneous populations of neurons ranging in scale from single regions to networks of activities across the whole brain \cite{deco2008dynamic,destexhe2009wilson,hopfield1982neural,breskin2006percolation,schneidman2006weak,tkacik2006ising,weistuch2020younger}.

But, there remain important macroscopic brain behaviors that cannot be captured by WC-like models \cite{muir1979simple,chow2019before}.  One example is theta oscillations in the hippocampus, which have multiple superimposed frequencies, and are thought to be critical for memory formation and storage \cite{buzsaki2002theta,buzsaki2004neuronal,colgin2013mechanisms}.  They are believed to be generated through recurrent feedback involving only excitatory neurons \cite{buzsaki2002theta}.  Another example is gamma oscillations, which are high-frequency chaotic firing patterns associated with a wide-range of complex brain activities \cite{buzsaki2012mechanisms}.  They are believed to arise in networks of inhibitory neurons.  In contrast, WC-like models achieve only simple oscillations, and require both excitatory and inhibitory neurons to achieve them.  Here, we describe a model of similar simplicity, but that also accounts for refractory periods of neurons, which we show to produce broader behaviors including complex oscillations \cite{muir1979simple,feldman1975large}.

\begin{figure*}
    \centering
    \includegraphics[width=.6\linewidth]{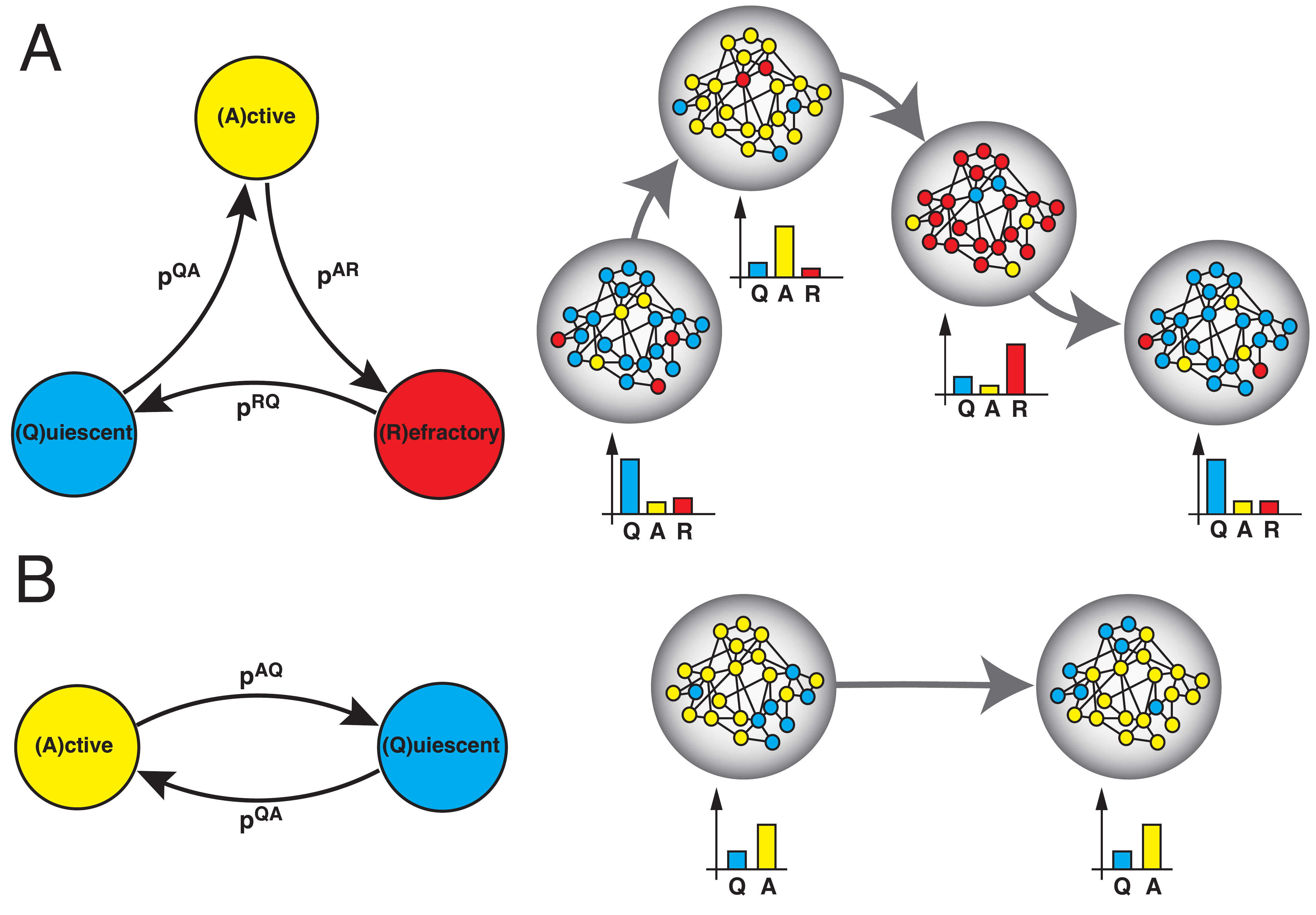}
    \caption{\footnotesize{{\bf Neural oscillations depend on the refractory period.}  (a).  A network representation of the model.  Left.  Neural activity represented as a Markov chain.  Here the rates ($p^{QA}$, $p^{AR}$, and $p^{RQ}$) determine the occupation probabilities of the three states: $Q$, $A$, and $R$.  Right.  The fraction of neurons in each state evolves over time.  Active neurons (yellow) can either excite (as shown) or inhibit neighboring quiescent neurons (blue) by modulating the average firing probability $p^{QA}$.  Complex oscillations emerge when these neurons must wait to fire again (red).  (b).  The Wilson-Cowan model cannot describe complex oscillations.  Without the refractory state (left), the fraction of neurons in each state does not evolve over time (right).}}  
    \label{fig:diagram}
\end{figure*}

The novelty of the present work is two-fold.  First, the model physics is different, and allows for a refractory period (see Fig. \ref{fig:diagram}a compared to the WC model in Fig. \ref{fig:diagram}b).  Second, we treat the stochastic dynamics of the model using {\it Maximum Caliber} (Max Cal), a principle of statistical inference that applies to systems of pathways and/or systems of dynamical processes, which draws more directly on data and is freer of unwarranted model assumptions  \cite{presse2013principles,dixit2018perspective,ghosh2020maximum,weistuch2020inferring}.

\subsection*{The physics of the model}
Here are the modeling details.  We represent a generic network of $N$ neurons (labeled $i=1,2,\hdots,N$) as a graph; nodes represent each neuron and edges are synaptic connections (Fig. \ref{fig:diagram}).  Each node ($i$) also has a time-dependent state $S_i(t)$, representing the activity of a neuron.  In particular, the nodes of our network can be in any one of three states: quiescent (Q), or silent but able to fire; active (A), or firing; refractory (R), or unable to fire.  Additionally, the states of each node evolve stochastically over time: $Q\to A \to R \to Q$.  The rate of each of these transitions is then chosen to reflect the biophysical dynamics of real neurons.  

We use the Principle of Maximum Caliber (Max Cal) to infer these transitions directly from the data \cite{presse2013principles,dixit2018perspective,ghosh2020maximum}.  Here Max Cal, the dynamical extension of Maximum Entropy, provides the simplest and least-biased model consistent with a few known transition rates, such as average neuronal firing rates and correlations \cite{tkacik2006ising,presse2013principles,dixit2018perspective,ghosh2020maximum}.  This model takes the form of a probability distribution $P$ over different dynamical trajectories $\Gamma$ of populations of neurons.

Using Max Cal, we model how the fraction of neurons in each state ($\pi_Q$, $\pi_A$, and $\pi_R$) evolve over time.  While our approach is applicable to any number of neurons, we focus on the case when this number is large.  We maintain our focus here for two reasons.  First, it presents an enormous simplification, as we can study the long-time behavior of our model using mean-field theory \cite{deco2008dynamic,omurtag2000simulation,gerstner2000population,jirsa1997derivation}.  Second, it is often a reasonable approximation, as system behaviors converge to their means when their number of components $N$ is large. 

\subsection*{Obtaining the stochastic dynamics of the model using Maximum Caliber}
Here we ask how simple neuronal interactions might give rise to complex patterns of brain dynamics.  To answer this, we use Max Cal to build a minimal model of neural dynamics.  Here, the Caliber $\mathcal{C}$ is defined as the path entropy over the probability distribution of trajectories $P_{\Gamma}$ subject to a prespecified set of constraints:
\begin{equation}
    \mathcal{C}[P_{\Gamma}]=-\sum_{\Gamma}P_{\Gamma}\log P_{\Gamma}+\sum_{i,\Gamma} \lambda_i A_{i,\Gamma}P_{\Gamma}
\end{equation}
where $\lambda_i$ are the Lagrange multipliers constraining generic average quantities $\langle A_i \rangle$.  Here the quantities that we measure are the transitions of nodes between different states: $l_i^{QA}(t)$, $l_i^{AR}(t)$, and $l_i^{RQ}(t)$.  In particular, $l_i^{QA}(t)$ is $1$ if the $i$th node transitions from Q to A during the time interval $[t,t+1]$ and is otherwise $0$; the other transition indicators are defined similarly.  We thus want to constrain our model in such a way to preserve the average transition rate between each pair of states:
\begin{align}
    r_{QA}=\frac{1}{N}\Big\langle \sum_{i=1}^N l_{i}^{QA}(t)\Big\rangle=\Big\langle \pi_Q(t)p^{QA}(t)\Big\rangle \nonumber\\
    r_{AR}=\frac{1}{N}\Big\langle \sum_{i=1}^N l_{i}^{AR}(t)\Big\rangle= \Big\langle \pi_A(t)p^{AR}(t)\Big\rangle \nonumber \\
    r_{RQ}=\frac{1}{N}\Big\langle \sum_{i=1}^N l_{i}^{RQ}(t)\Big\rangle= \Big\langle \pi_R(t)p^{RQ}(t)\Big\rangle
    \label{eq:means}
\end{align}
Here $\langle \circ \rangle$ denotes an average over time and the second set of equalities hold when the number of neurons $N$ is large.  The average rates $r_{QA}$, $r_{AR}$, and $r_{RQ}$ are computed from experimental data as the time-averaged fraction of nodes transitioning from $Q\to A$, $A\to R$, and $R\to Q$ respectively.  In contrast, the right-hand sides of the above equations are computed over the different trajectories that our inferred model will produce.  Here these averages are constrained using the Lagrange multipliers $h^{QA}$, $h^{AR}$, and $h^{RQ}$ respectively (see Appendices A and B).  

These Lagrange multipliers can then be incorporated into the transition probabilities $p^{QA}$, $p^{AR}$, and $p^{RQ}$ as discussed in Appendix B.  Here $p^{AR}$ and $p^{RQ}$ are constants and are functions of their respective Lagrange multipliers.  More directly, $p^{AR}$ (resp. $p^{RQ}$) can be computed as the average fraction of refracting $A$ (resp. quiescing $R$) per unit time.  In contrast, a key property of neurons is their ability to communicate by altering the firing activity of their neighbors.  Specifically, firing neurons can either increase (excite) or decrease (inhibit) the probability that other quiescent neurons fire.  Here we include this with the additional constraint:
\begin{equation}
    C=\frac{1}{N}\Big\langle \sum_{i=1}^N l_{i}^{QA}(t)N_A(t)\Big\rangle=N\Big\langle \pi_Q(t)\pi_A(t)p^{QA}(t)\Big\rangle
    \label{eq:correlation}
\end{equation}
Where $N_A(t)$ is the number of active neurons at time $t$.  A positive value of $C$ thus represents a population of excitatory neurons, as the firing probability of additional nodes increases with the number of currently active nodes.  Conversely, a negative value of $C$ represents an inhibitory population, whereby the activation of a few nodes suppresses subsequent firing of additional nodes.  This constraint is enforced by the Lagrange multiplier $J$, the coupling constant.  Thus, the transition probability $p^{QA}$ is a function of both the raw firing probability of a neuron (controlled by $h=h^{QA}$) and the feedback strength, $J$.  This relationship is given by (see Appendix \ref{sec:MF}):
\begin{equation}
    p^{QA}(\pi_A)=\frac{e^{h+J\pi_A}}{1+e^{h+J\pi_A}}
    \label{eq:pnew}
\end{equation}

And thus $h$ and $J$ can be alternatively computed by fitting the shape of $p^{QA}$ for different values of $\pi_A$.  Taken together, our model is a function of 4 parameters ($p^{AR}$, $p^{RQ}$, $h$, and $J$) each uniquely chosen to reproduce our 4 experimental constraints.

\subsubsection*{The mean-field equations}
Here we compute the time-evolution of the average fraction of neurons in each state ($\pi_Q$, $\pi_A$, and $\pi_R$).  Before proceeding, we make a few notational simplifications to enhance readability.  First, we use $\Delta$ to refer to the change in a variable over time.  For example, $\Delta \pi_A(t)=\pi_A(t+1)-\pi_A(t)$.  And second, aside from their initial definitions, we implicitly assume the time dependence of our variables and drop $(t)$ when writing our equations.  For example, $\pi_A(t)$ will just be written as $\pi_A$.  And $\Delta \pi_A(t)$ will just be $\Delta \pi_A$.  Thus, after maximizing the caliber subject to our four constraints and computing the average (mean-field) dynamics (see Appendices \ref{sec:MaxCal} and \ref{sec:MF}), we find that our system can be described by two coupled equations:
\begin{align}
    &\Delta \pi_Q=(1-\pi_A-\pi_Q)p^{RQ}-\pi_Qp^{QA} \nonumber \\
    &\Delta \pi_A=\pi_Qp^{QA}-\pi_Ap^{AR} \label{eq:mfe}
\end{align}
Here we have eliminated the corresponding third equation for $\Delta \pi_R$ using the constraint that the fractions of nodes of each type sum to unity (i.e. the number of neurons is fixed).

In contrast to typical modeling approaches, we have made no assumptions in deriving these equations other than the fact that our experimentally observed constraints are reasonably descriptive of neural dynamics.  Thus, we expect our model to be widely-applicable, even when other previous models fail.

Also, each of our parameters has a clear biological interpretation.  First, $p^{AR}$ and $p^{RQ}$ control the average amount of time neurons spend (respectively) active and refractory.  Thus when $p^{AR}$ is large (as might be expected of real neurons), nodes are only briefly active.  On the other hand, $p^{RQ}$ might be expected to be small, as biological neural oscillation occurs at a relatively low frequency (an action potential lasts 1 ms, but the fastest oscillations have a period of about 10 ms).  Reflecting these requirements, we fix $p^{AR}$ and $p^{RQ}$ at $0.8$ and $0.01$ respectively.  Additionally, $h$, the unit-less {\it average firing threshold}, controls the fraction of neurons that fire spontaneously.  Thus we should have $h<0$, reflecting a low-level of baseline activity.  Finally, $J$ reflects feedback, or {\it synaptic coupling}, between neighboring neurons and can be either positive (excitatory) or negative (inhibitory).

We study two general classes of brain oscillations, corresponding to the network activities of excitatory ($J>0$) and inhibitory ($J<0$) neurons.  Here, excitatory oscillations are characterized by high amplitude waves of activity followed by long periods of silence during which most neurons are refractory \cite{buzsaki2002theta,buzsaki2004neuronal}.  In contrast, networks of inhibitory neurons fire asynchronously, producing low amplitude, high-frequency oscillations \cite{buzsaki2012mechanisms,brunel2008sparsely,korn2003there}.  And unlike WC, both of these behaviors can be exhibited by our model \cite{muir1979simple,feldman1975large}.

\subsection*{Model properties}

The formulation of WC-like models is based on quasi-steady state dynamics and can only be applied to simple systems of neurons, oscillating at a single frequency (see Appendix \ref{sec:WC} and \cite{muir1979simple,feldman1975large}).  In contrast, the behaviors of real neurons are more complex and have been difficult to describe mechanistically \cite{buzsaki2004neuronal,chow2019before}.  We next demonstrate the significant improvements of our model over these previous approaches.

\begin{figure*}[t!]
    \centering
    \includegraphics[width=.8\linewidth]{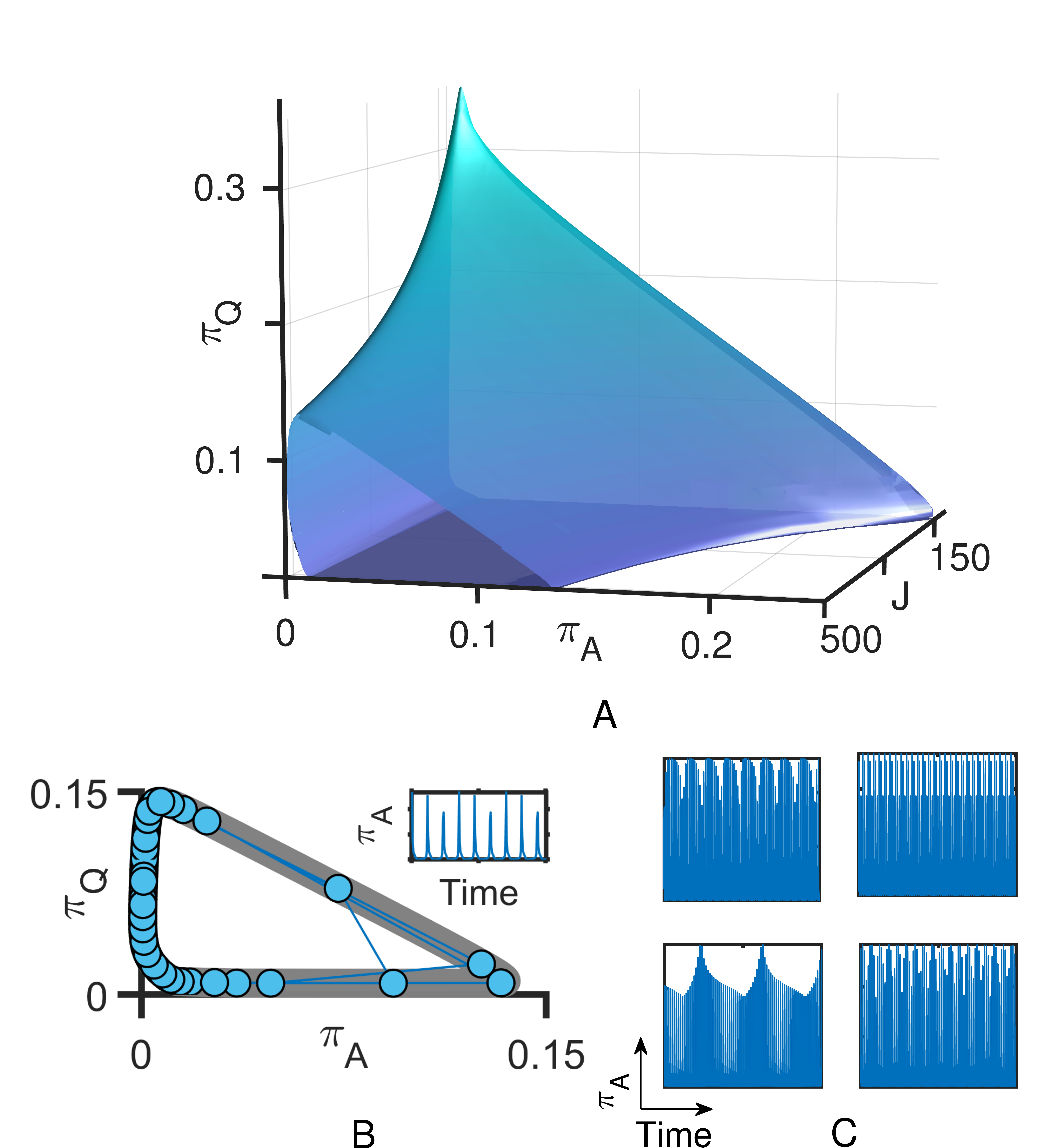}
    \caption{\footnotesize{{\bf Excitatory couplings produce complex oscillations}.
 (a).  The phase plane ($\pi_Q$ vs $\pi_A$) for different values of $J$ (at $h=-5$), illustrating the emergence of oscillations (rings).  (b).  A typical trajectory (blue) in phase space (grey) and over time (inset).  Because $\pi_Q$ and $\pi_A$ vary slightly with each cycle, the oscillatory amplitude changes over time.  These changes are very sensitive to $J$.  (c).  Examples of different oscillatory patterns for different values of $J$.}}
    \label{fig:osc}
\end{figure*}

Groups of excitatory neurons tend to fire in synchronized bursts called {\it avalanches}.  The resulting oscillations are characterized by high amplitude waves of activity, with multiple frequencies, followed by long periods of silence during which most neurons are refractory \cite{buzsaki2004neuronal}.  Unlike WC, our model explains how these patterns might emerge (Fig. \ref{fig:osc}.  In particular, Fig. \ref{fig:osc}a, depicting the phase plane of our model, shows the emergence of oscillatory activity (rings) when the coupling $J>0$ is nestled within a critical region.  Here the amplitude of each oscillation can vary with every cycle (Fig. \ref{fig:osc}b), producing the multi-frequency bands expected of real neurons \cite{buzsaki2002theta,vladimirski2008episodic,bacak2016mixed}.  Unlike WC, by only slightly tuning $J$, our model predicts the emergence of highly distinct patterns of activity (Fig. \ref{fig:osc} c).  And indeed, a similar mechanism is thought to underlie tremendous information capacity of real networks of neurons \cite{panzeri2015neural,averbeck2006neural,vladimirski2008episodic}.

\begin{figure*}
    \centering
    \includegraphics[width=.8\linewidth]{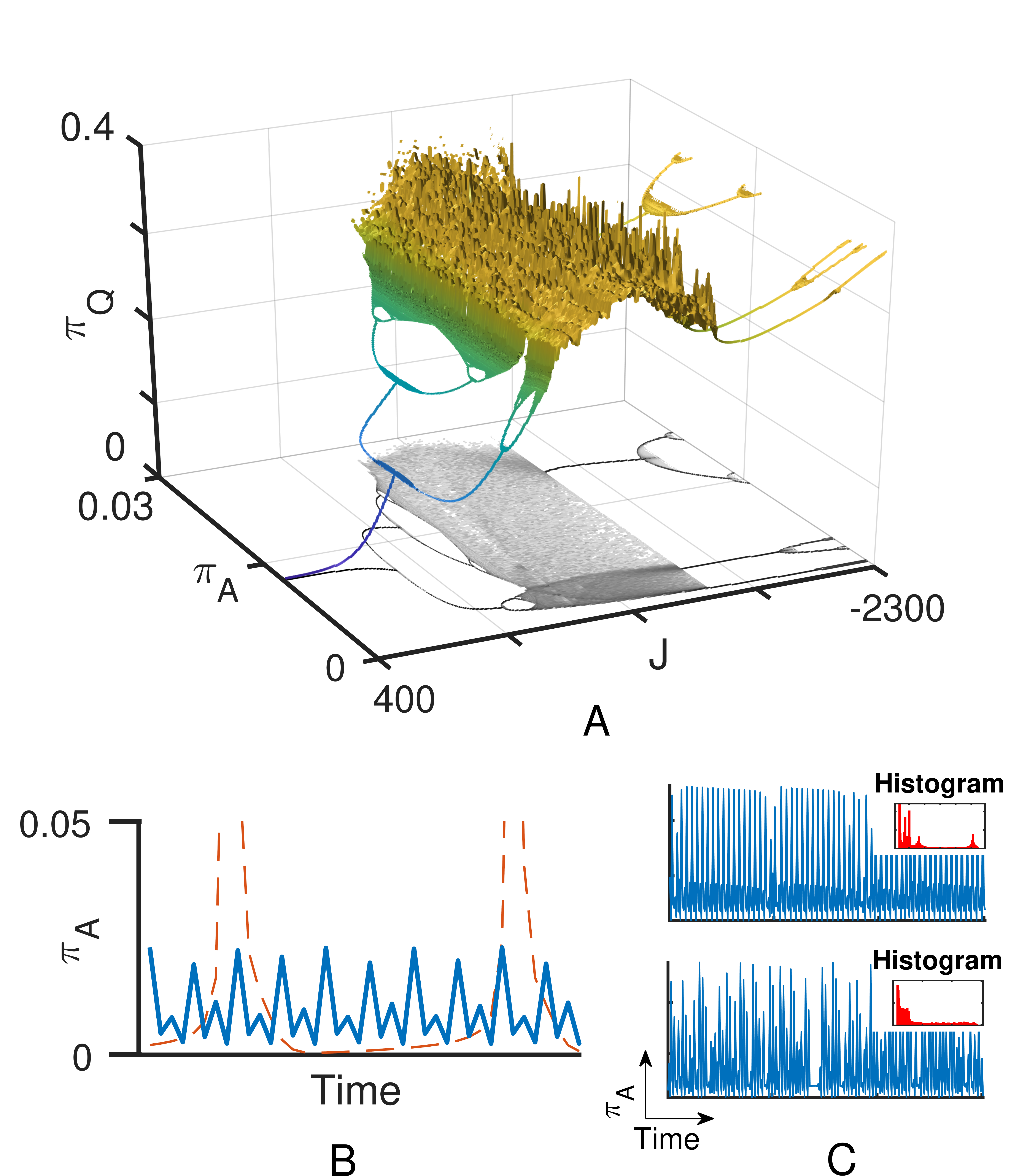}
    \caption{\footnotesize{{\bf Inhibitory couplings produce chaotic oscillations}.  (a).  The phase plane ($\pi_Q$ vs $\pi_A$) and its projection (black) for different values of $J$ (at $h=-1$)   The number of points (for each $J$) corresponds to the period of the associated oscillation.  As $J$ is decreased, the oscillations become chaotic and aperiodic (orange).  (b).  Comparison of inhibitory (blue) to excitatory (orange) oscillations produced by our model.  (c).  Examples of different chaotic oscillatory patterns along with a histogram of $\pi_A$ over time (inset).  Here information is stored, not in the timing, but in the probabilities of different amplitudes.}}
    \label{fig:chaos}
\end{figure*}

\clearpage

At the other extreme, recurrent inhibitory networks of neurons have been shown to produce high-frequency and sometimes chaotic firing patterns \cite{buzsaki2012mechanisms,korn2003there}.  In contrast to excitatory networks, inhibitory neurons fire in small bands of only a few neurons at a time.  As the strength of this inhibition is increased, these neurons fire asynchronously and chaotically \cite{buzsaki2012mechanisms}.  Fig. \ref{fig:chaos} describes how these features emerge from our model.  Here, Fig. \ref{fig:chaos}a. depicts the phase plane of our model for different values of $J$ ($J<0$).  The number of points for each $J$ corresponds to the period of the inhibitory oscillations.  As $J$ is decreased, this period continually doubles until it diverges to infinity and chaos emerges.  Because inhibitory neurons fire as far apart as possible, they oscillate with a much higher frequency (as well as a lower amplitude) as compared to excitatory neurons (Fig. \ref{fig:chaos}b) \cite{buzsaki2004neuronal,buzsaki2012mechanisms}.  

And despite appearing to have almost noise-like dynamics, these chaotic firing patterns robustly store information in their probability distributions of amplitudes (Fig. \ref{fig:chaos}c and inset).  And thus, the asynchronous oscillations in real networks of inhibitory neurons can be seen as information transmission that is fast and robust to noise \cite{buzsaki2012mechanisms,brunel2008sparsely}.  Also, hidden within the chaotic region are occasional windows of stable oscillations that are seen when $J$ is very negative (Fig. \ref{fig:chaos}a).  Whether pathological or strategic, this suggests that real networks of neurons may be able to flexibly switch between qualitatively different patterns of firing activity by only slightly changing their synaptic coupling \cite{korn2003there}. 

\begin{figure}[h!]
    \centering
    \includegraphics[width=\linewidth]{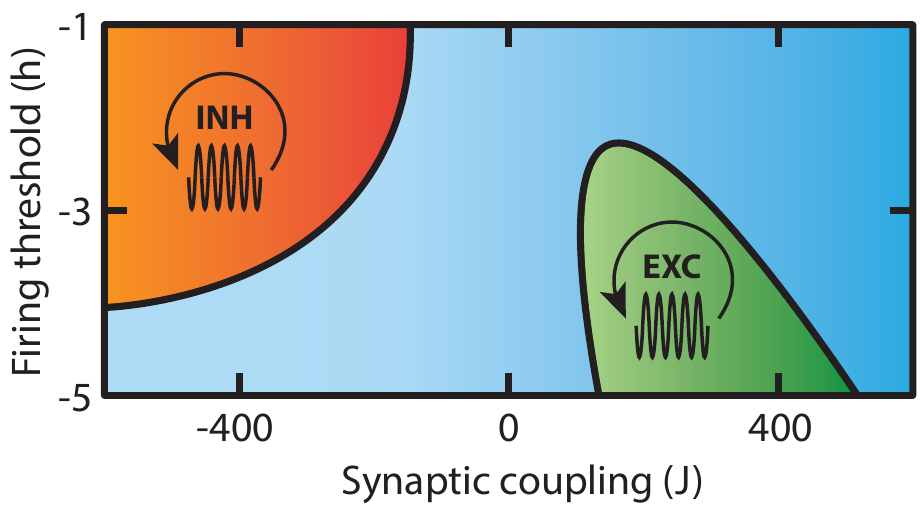}
    \caption{\footnotesize{{\bf The phase diagram of our model, depicting the emergence of excitatory (green) and inhibitory (orange) oscillations.}  In the blue region, brain activity is constant over time.  In contrast to WC-like models, however, oscillations can be produced by tuning $h$ and $J$.}}
    \label{fig:phase-diagram}
\end{figure}

Taken together, the general behavior of our model changes dramatically, in biologically expected ways, as its parameters are varied.  These findings are summarized in Fig. \ref{fig:phase-diagram}, illustrating how these behaviors change with $h$ and $J$ (with $p^{RQ}$ and $p^{AR}$ fixed at their previous, biologically plausible values).  In particular, as long as $p^{RQ}$ and $p^{AR}$ are biologically appropriate, our model exhibits roughly three different behaviors (corresponding to the three colors in \ref{fig:phase-diagram}): constant (equilibrium) activity and both excitatory and inhibitory oscillations (including chaos).  In contrast, the Wilson-Cowan model only exhibits the former behavior (Appendix \ref{sec:WC}).  And analysis of the locations and properties of each of these regimes can be easily performed using only standard techniques (Appendix \ref{sec:Bifurcation}).  Thus our model explains a huge variety of complex, natural phenomena in a simple and practical way.  In particular, $h$ (i.e. the mean firing probability) can be manipulated experimentally by applying an external voltage to a group of neurons \cite{breskin2006percolation,eckmann2007physics}.  Also, synaptic activity ($J$) can be manipulated \cite{breskin2006percolation}.  The predictions of our model (and even the phase diagram itself) can be easily tested experimentally. 

\section*{Discussion}

Here we have presented a new treatment of collective neural dynamics that starts from  only the most elementary biophysical of neurons, and basic stochastic dynamics.  We find a broad range of behaviors, even in the simplest case of only a single type of neuron (either excitatory or inhibitory).

Of course, many situations involve both types of neurons.  Nevertheless, there are some situations only involving a single type.  For example, theta-wave neuronal oscillations in the hippocampus are thought to play a considerable role in memory formation and spatial navigation \cite{buzsaki2002theta,colgin2013mechanisms}.  The currents driving these oscillations are believed to be primarily generated by recurrent excitatory-excitatory connections within the CA3 region of the hippocampus, whereby these neurons robustly synchronize using a ``relaxation'' mechanism akin to our model's predictions \cite{buzsaki2002theta,buzsaki2004neuronal}.  The present model suggests how these neurons can so easily toggle between and store the large number of complex oscillatory patterns required for their proper function \cite{buzsaki2004neuronal,llinas1988intrinsic,hutcheon2000resonance}.

Similarly, the emergence of chaotic neural dynamics has been seen experimentally and is believed to underlie high-frequency, gamma-band oscillations across the brain \cite{korn2003there,aihara1990chaotic,brunel2008sparsely}.  Our model generates these patterns with just inhibitory neurons \cite{buzsaki2012mechanisms}.  And, while chaotic dynamics might seem counterproductive for the brain, it has been theorized that these patterns are critical for information storage \cite{brunel2008sparsely,aihara1990chaotic,korn2003there,jia2012dynamics}.  And perhaps fluctuations into the occasional window of stability within this chaos play a role in pathologies such as epilepsy \cite{sato2017spatiotemporal}.

Our model is readily extended beyond a single type of neuron.  In particular, the Hopfield model of associative learning has been an essential starting point for much of the recent development in artificial neural networks \cite{hopfield1982neural,hopfield1986computing}.  In that case, each pair of neurons is assigned its own learned coupling $J_{ij}$, representing the storage of unique patterns of activity.  The present model may allow generalization beyond the Hopfield model \cite{destexhe2009wilson}.

\begin{acknowledgments}
The research was funded by the WM Keck Foundation (LRMP, KD), the NSF BRAIN Initiative (LRMP, KD: ECCS1533257, NCS-FR 1926781), and the Stony Brook University Laufer Center for Physical and Quantitative Biology (KD).
\end{acknowledgments}

\section*{COMPETING INTERESTS}
The authors declare no competing financial interests.

\bibliography{neuron_refs.bib}

\bibliographystyle{unsrt}
\onecolumngrid

\appendix

\section{Maximizing caliber for Markovian processes}\label{sec:MaxCal}
Here we summarize how to apply Max Cal to Markovian systems.  Here the trajectories $\Gamma$ of some variable $S$ are defined as: $\Gamma=\{S_0,S_1,\hdots, S_T\}$.  Our goal is to infer $P_\Gamma$ using some given information, or constraints.  First, since the process is Markovian:
\begin{equation}
    P_\Gamma=\pi(S_0)\prod_{t=1}^T P(S_{t}|S_{t-1})
\end{equation}
where the vertical bar is used to denote the conditional probability.  Here $P$ denotes the transition probabilities and $\pi$ denotes a distribution over states.  In particular, if the Markov chain is allowed to reach a steady-state distribution $\pi$:
\begin{equation}
    \pi(S_t)=\sum_{S_{t-1}} P(S_t|S_{t-1})\pi(S_{t-1}) ,\qquad \sum_{S_t}\pi(S_t)=1
\end{equation}

The path entropy $\mathcal{E}$ can then be written as:
\begin{equation}
    \mathcal{E}\{P\}=-\sum_{\Gamma}P_\Gamma\log P_\Gamma=\\
    -\sum_{S_0}\pi(S_0)\log \pi(S_0)-\sum_{t=1}^T\sum_{S_t}\sum_{S_{t-1}}\pi(S_{t-1})P(S_t|S_{t-1})\log P(S_t|S_{t-1})
\end{equation}
which for $T$ large reduces to:
\begin{equation}
    \frac{1}{T}\mathcal{E}\{P\}=-\sum_{S_a}\sum_{S_b}\pi(S_a)P(S_b|S_a)\log P(S_b|S_a)
\end{equation}
for generic subsequent times $a$ and $b$.  We now write our caliber $\mathcal{C}$ as:
\begin{equation}
    \frac{1}{T}\mathcal{C}\{P\}=\sum_{S_a}\pi(S_a)\sum_{S_b}P(S_b|S_a)\Bigg[-\log P(S_b|S_a)+ \mu(S_a) +\sum_i \lambda_i A_i(S_a,S_b)\Bigg]
    \label{eq:cal}
\end{equation}
Here $\mu(S_a)$ ensures that the transition probabilities $P(S_b|S_a)$ are properly normalized (sum to $1$).  Additionally the Lagrange multipliers $\lambda_i$ enforce the constraints of 
$\langle A_i(S_a,S_b)\rangle$, such as the mean transition rates discussed in the main text.  We find the trajectory distribution that maximizes the caliber $\mathcal{C}$:
\begin{equation}
    \frac{\partial \mathcal{C}}{\partial P(S_b|S_a)}=0 \implies -\log P(S_b|S_a) +\mu(S_a) +\sum_i \lambda_i A_i(S_a,S_b)-1=0
\end{equation}
Therefore:
\begin{equation}
    P(S_b|S_a)=e^{\mu(S_a)-1+\sum_i \lambda_i A_i(S_a,S_b)}
\end{equation}
Since the distributions need to be normalized, we have that:
\begin{equation}
    P(S_b|S_a)=\frac{e^{\sum_i \lambda_i A_i(S_a,S_b)}}{\sum_{S_b}e^{\sum_i \lambda_i A_i(S_a,S_b)}}
    \label{eq:MaxCal}
\end{equation}
Finally, using our original constraints (Eq. \ref{eq:cal}), we can uniquely determine the Lagrange multipliers $\lambda_i$.

\section{Deriving the mean-field model from Max Cal}\label{sec:MF}
Here our goal is to understand how the constraints Eqs. \ref{eq:means} and \ref{eq:correlation} give rise to our mean-field model Eq. \ref{eq:mfe}.  First, we use $S(t)=\{S_1(t),S_2(t),\hdots,S_N(t)\}$ to denote the states of all nodes at time $t$.  Second, the number of nodes in each state are then given (respectively) by $N_Q(t)$, $N_A(t)$, and $N_R(t)$.  And finally, transitions are indicated by the functions $l_i^{QA}(t)$, $l_i^{AR}(t)$, and $l_i^{RQ}(t)$ as indicated in the main text.  We next follow the general procedure laid out in Appendix \ref{sec:MaxCal} (Eq. \ref{eq:MaxCal}) to infer the transition probabilities $P\Big(l_i^{QA}|S\Big)$, $P\Big(l_i^{AR}|S\Big)$, and $P\Big(l_i^{RQ}|S\Big)$.  In particular, each quiescent (Q) node fires ($Q\to A$) with probability:
\begin{equation}
    P\Big(l_i^{QA}|S\Big)=\frac{e^{h^{QA}+J^{\ast}N_A}}{1+e^{h^{QA}+J^{\ast}N_A}}=p^{QA}
    \label{eq:QA}
\end{equation}

Similarly, each active (A) node becomes refractory ($A\to R$) with probability:
\begin{equation}
    P\Big(l_i^{AR}|S\Big)=\frac{e^{h^{AR}}}{1+e^{h^{AR}}}=p^{AR}
    \label{eq:AR}
\end{equation}
and each refractory (R) node quiesces ($R \to Q$) with probability:
\begin{equation}
    P\Big(l_i^{RQ}|S\Big)=\frac{e^{h^{RQ}}}{1+e^{h^{RQ}}}=p^{RQ}
    \label{eq:RQ}
\end{equation}

Eqs. \ref{eq:QA}, \ref{eq:AR}, and \ref{eq:RQ} provide the rules by which our simple network of neurons evolves over time.  However, here we are primarily interested in how the population dynamics of a group of neurons changes over time, in particular $N_Q$, $N_A$, and $N_R$.  For example, changes in $N_Q$ can occur in two different ways.  First, nodes in $R$ can quiesce ($R\to Q$), adding to the total number of $Q$ nodes.  Second, nodes in $Q$ can fire ($Q\to A$), subtracting from the total number of $Q$ nodes.  Here we denote the number of each kind of transition as $N_{RQ}$, $N_{QA}$, and $N_{AR}$.  The number of nodes of each type at time $t+1$ is then given by:
\begin{align}
    \Delta N_Q=N_{RQ}-N_{QA} \nonumber\\
    \Delta N_A=N_{QA}-N_{AR} \nonumber\\
    \Delta N_R=N_{AR}-N_{RQ} \label{eq:dyn}
\end{align}

In reality though, we only have two dynamical equations since:
\begin{equation}
    N_Q+N_A+N_R=N
    \label{eq:consprob}
\end{equation}
for all $t$.  Additionally, since each transition is independent, the number of transitions of each type is binomially-distributed:
\begin{align}
N_{RQ}\sim B(N_R,p^{RQ}) \nonumber\\
N_{QA}\sim B(N_Q,p^{QA}) \nonumber\\
N_{AR}\sim B(N_A,p^{AR})
\end{align}
Here we use $B(N,p)$ as shorthand for the two-parameter binomial distribution; $N$ is the number of trials and $p$ is the probability of each success (here a transition of a particular node).  We next ask how simple neuronal interactions might give rise to complex patterns of brain dynamics.  In particular, we use {\it mean-field theory} to explore how our previous equations behave when the number of neurons is large \cite{deco2008dynamic}.  To simplify our analysis, we divide Eq. \ref{eq:dyn} by the number of nodes $N$ and instead follow how the average fraction of nodes in each state ($\pi_Q$, $\pi_A$, and $\pi_R$) change over time.  Since the mean of a binomially-distributed random variable $B(N,p)$ is $Np$, the average dynamics of our model are given by:

\begin{align}
    \Delta \pi_Q=\pi_Rp^{RQ}-\pi_Qp^{QA}\nonumber\\
    \Delta \pi_A=\pi_Qp^{QA}-\pi_Ap^{AR}\nonumber\\
    \Delta \pi_R=\pi_Ap^{AR}-\pi_Rp^{RQ}\label{eq:mf}
\end{align}

But, we can eliminate contributions from $\pi_R$ using Eq. \ref{eq:consprob}.  In addition, to keep all variables in terms of the fractions $\pi$, we define $J=J^{\ast}N$ and thus arrive at our final relationships Eqs. \ref{eq:mfe} and \ref{eq:pnew}.

\section{Deriving the Wilson-Cowan Model from Max Cal}\label{sec:WC}
Here we show how the widely-used Wilson-Cowan model emerges as a special case of our more general Max Cal model.  For simplicity, we focus on only a single type of neuron, but the derivation (as well as our model) can almost trivially be extended to any number of neural types by adding additional couplings.  Here we start from our mean-field model Eq. \ref{eq:mfe}.  One implicit assumption of the Wilson-Cowan model is that the number of refractory neurons ($\pi_R=1-\pi_A-\pi_Q$) is in a quasi-steady state $\Delta \pi_R \approx 0$ \cite{chow2019before}.  Thus adding together both parts of Eq. \ref{eq:mfe},
\begin{equation}
    (1-\pi_A-\pi_Q)p^{RQ}-\pi_Ap^{AR}=0\implies \pi_Q=1-\pi_A\Bigg(1+\frac{p^{AR}}{p^{RQ}}\Bigg)
\end{equation}

Next we define the constant $r=1+\frac{p^{AR}}{p^{RQ}}$.  Substituting this back into our equation for $\Delta \pi_A$:
\begin{equation}
    \Delta \pi_A=p^{QA}(1-r\pi_A)-\pi_Ap^{AR}
    \label{eq:wilsco}
\end{equation}
Now defining $p^{AR}=1/\tau$ and rearranging, our equation turns into the exact same form as that from the Wilson-Cowan equation \cite{wilson1972excitatory,chow2019before}.  This re-derivation also tells us exactly when Wilson-Cowan breaks.  In particular, the quasi-steady-state assumption is false when the rate of change of recovering neurons is large.  In other words, the Wilson-Cowan model cannot describe strongly-coupled behaviors such as avalanches and intrinsic oscillations.  In contrast, our Max Cal model provides a much more complete picture of neural dynamics while retaining the simplicity of the original Wilson-Cowan model.

\section{Bifurcation analysis}\label{sec:Bifurcation}
Here we use local stability analysis to explore how our model transitions between simple equilibrium behavior and complex oscillatory dynamics as its parameters are varied.  To achieve this, we compute the equilibrium state of our model and ask how typical trajectories behave in its vicinity.  In general, a system is in equilibrium if it does not change over time.  Thus, the equilibrium states of our model are the coordinates where the LHS of Eq. \ref{eq:mfe} is $0$.  After standard algebraic manipulation, we find that the equilibrium point satisfies:
\begin{equation}
    \pi_A^{\ast}=\frac{p^{RQ}p^{QA}}{p^{RQ}p^{QA}+p^{QA}p^{AR}+p^{AR}p^{RQ}}=\frac{p^{RQ}p^{QA}}{p_D}, \qquad \pi_Q^{\ast}=\pi_A^{\ast}\frac{p^{AR}}{p^{QA}}
    \label{eq:equil}
\end{equation}

The behavior of trajectories near this point is then determined by the Jacobian matrix of derivatives $\bm{\mathcal{J}}=\frac{\partial (\pi_Q',\pi_A')}{\partial (\pi_Q,\pi_A)}$.  Here we use $'$ to denote a subsequent time step ($t+1$) and bold to denote matrices.  For Eq. \ref{eq:mfe}, the Jacobian is given by:
\begin{equation}
    \bm{\mathcal{J}}(\pi_Q,\pi_A)=
    \begin{pmatrix}
    1-p^{RQ}-p^{QA} & -p^{RQ}-M\\
    p^{QA} & 1-p^{AR}+M)
    \end{pmatrix}
    \label{eq:Jac}
\end{equation}
Here $M= \pi_QJp^{QA}(1-p^{QA})$.  To describe the stability, we must compute the eigenvalues of this matrix evaluated at the equilibrium $(\pi_Q^{\ast},\pi_A^{\ast})$.  In particular, when the magnitude of these eigenvalues (whether real or complex) are both less than 1, all trajectories rapidly approach the equilibrium point ($\pi_Q^{\ast},\pi_A^{\ast}$), i.e. the dynamics are stable.  But when one (or both) of these eigenvalues has magnitude greater than $1$, trajectories never reach equilibrium (the dynamics are unstable).  Additionally, this transition can occur in several different ways, giving rise to the different types of oscillations we observe.  In particular, excitatory oscillations occur when the real part of this eigenvalue is positive (leading to large oscillations between high $\pi_A$ and high $\pi_Q$).  In contrast, when the real part of this eigenvalue is negative, high-frequency, inhibitory oscillations occur.  To determine when oscillatory behaviors occur, we thus need to determine when the eigenvalues of $\bm{\mathcal{J}}$ change their stability.  To simplify the expression of these eigenvalues, we define:
\begin{equation}
    F=\frac{p^{RQ}+p^{QA}+p^{AR}-M}{2}
\end{equation}
The eigenvalues, $\lambda$, are then given by:
\begin{equation}
   \lambda= 1-F \pm \sqrt{F^2-p_D+p^{RQ}M}
   \label{eq:eig}
\end{equation}

We now have 3 cases to consider.  First, when the unstable eigenvalue is $1$ (excitatory).  Second, when it is $-1$ (inhibitory).  And third, when it is complex with $|\lambda|=1$ (either excitatory or inhibitory).

For the first case, we set $\lambda=1$ and solve Eq. \ref{eq:eig} to find the critical point $J_c$:

\begin{equation}
    J_c=-\frac{1}{\pi_A^{\ast}}W\Bigg(\frac{-e^{h}}{p^{AR}\pi_A^{\ast}}\Bigg)
\end{equation}
where $W(x)$ is the multi-valued Lambert W function.  For the $\lambda=-1$ (inhibitory case), $p^{QA}$ and $p^{RQ}$ are both expected to be small.  Solving Eq. \ref{eq:eig} after this approximation produces $J_c$:
\begin{equation}
    J_c=\frac{p^{RQ}+p^{AR}-2}{\pi_A^{\ast}p^{AR}}-\frac{1}{\pi_A^{\ast}}W\Bigg(-\frac{p^{RQ}+p^{AR}-1}{p^{AR}}e^{h+\frac{p^{RQ}+p^{AR}-2}{p^{AR}}}\Bigg)
\end{equation}

And finally, solving the complex case exactly produces $J_c$:
\begin{equation}
    \frac{p^{RQ}+p^{AR}}{(1-p^{RQ})\pi_A^{\ast}p^{AR}}-\frac{1}{\pi_A^{\ast}}W\Bigg(-\frac{p^{RQ}+p^{AR}+1-\frac{p^{RQ}}{\pi_A^{\ast}}}{(1-p^{RQ})p^{AR}}e^{h+\frac{p^{RQ}+p^{AR}}{(1-p^{RQ})p^{AR}}}\Bigg)
\end{equation}

Additionally, if $p^{RQ} \ll p^{QA}$ (which is almost always the case for biologically plausible sets of parameters):
\begin{equation}
\pi_A^{\ast}\approx \frac{p^{RQ}}{p^{RQ}+p^{AR}} 
\end{equation}

Thus, the above $3$ scenarios describe three sets of critical transitions between different types of oscillations.  The first ($\lambda=1$) and last ($\lambda$ complex) both exclusively correspond to the emergence of excitatory oscillations.  In contrast, the second case ($\lambda=-1$) corresponds to the emergence of inhibitory oscillations.  Most importantly, the Lambert function $W(x)$ is only defined when $x\geq e^{-1}$.  And, when $-e^{-1}<x<0$, the Lambert function has two solutions (corresponding to the beginning and end of oscillatory behavior).  Thus we have found an analytical relationship between our model parameters and the emergence of qualitatively distinct biological patterns.

\end{document}